\documentclass[twocolumn,aps,pra,floatfix,longbibliography]{revtex4-2}

\usepackage{amsmath}
\usepackage{txfonts}
\usepackage{microtype}
\usepackage{graphicx}
\usepackage{color}
\usepackage{ulem}
\usepackage{bm}

\begin{document}

\title{Relativistic analytical R-matrix (ARM) theory for strong-field ionization}

\author{Michael Klaiber}\email{klaiber@mpi-hd.mpg.de}
\affiliation{Max-Planck-Institut f\"ur Kernphysik, Saupfercheckweg 1, 69117 Heidelberg, Germany}
\author{Karen Z. Hatsagortsyan}\email{k.hatsagortsyan@mpi-hd.mpg.de}
\affiliation{Max-Planck-Institut f\"ur Kernphysik, Saupfercheckweg 1, 69117 Heidelberg, Germany}
\author{Christoph H. Keitel}
\affiliation{Max-Planck-Institut f\"ur Kernphysik, Saupfercheckweg 1, 69117 Heidelberg, Germany}

\date{\today}

\begin{abstract}

The analytical R-matrix (ARM) theory has been known for an efficient description of the Coulomb effects of the atomic core in strong-field ionization in the nonrelativistic regime. We generalize the ARM theory into the relativistic domain aiming at the application to strong-field ionization of highly-charged ions in ultrastrong laser fields. Comparison with the relativistic Coulomb-corrected strong field approximations (SFA) is provided, highlighting the advantages and disadvantages. The weakly relativistic asymptotics and its accordance with the nondipole Coulomb-corrected SFA are examined. As an example of a physical application of the relativistic ARM, the  Coulomb enhancement of tunneling ionization probability for highly-charged ions at the cutoff of the direct channel is discussed.

\end{abstract}

\maketitle

\section{Introduction}

Advances in the experimental technique for high-resolution measurements of the photoelectron and ion momentum distributions \cite{COLTRIM,Weger_2013} allowed the recent extension of experimental investigations of strong-field ionization beyond the dipole regime \cite{Smeenk_2011,Ludwig_2014,Maurer_2018,Willenberg_2019,Hartung_2019,Haram_2019,Haram_2020,Hartung_2021,
 Lin_2021,Lin_2022,Lin_2022a}. The leading nondipole effect is the radiation pressure which is responsible for  the partitioning of the absorbed photon momentum between the photoelectron and the parent ion in  strong-field ionization \cite{Smeenk_2011,Klaiber_2013,Chelkowski_2014,Cricchio_2015,Bandrauk_2015,He_2017,He_2022}. Interesting dynamical properties arise due to the interplay between  Coulomb effects of the atomic core and the nondipole effects \cite{Forre_2006,Ludwig_2014,Keil_2017,Tao_2017,Maurer_2018,Danek_2018b,Danek_2018,Danek-II_2018,
 Danek_2019,Willenberg_2019}.
The nondipole theory has been developed for the interpretation of experimental results, including the strong field approximation (SFA) \cite{Klaiber_2005,Klaiber_2013,He_2017,He_2022,He_2022a,Klaiber_2022,Ni_2020,Habibovic_2022,
Lund_2021,Boening_2021,Madsen_2022,Fritzsche_2022,Kahvedzic_2022,Mao_2022}, the numerical solution of the time-dependent Schr\"odinger equation (TDSE) \cite{Ni_2020,Brennecke_2018,Brennecke_2021}, as well as  the classical trajectory Monte Carlo (CTMC) simulations \cite{Klaiber_2017,Jones_2021,Kelley_2021,Lin_2022}.

While presently strong laser fields up to the intensity of $10^{23}$~W/cm$^2$ are achievable \cite{Yoon_2021}, the relativistic regime of the laser-atom interaction in ultrastrong fields is far from deep experimental scrutiny. This is because the most interesting dynamics, including electron correlations, is expected when the atomic and laser fields are of the same magnitude. The latter necessitates dealing with an atomic system of  highly-charged ions (HCI), which are extremely difficult to handle experimentally. The pioneering experiment in this field by Moore \textit{et al.} \cite{Moore_1999} more than 20 years ago at an intensity of $3\times 10^{18}$~W/cm$^2$, has been followed by a series of fine experiments aimed at the observation of signatures of the atomic bound dynamics in the photoelectron momentum distribution (PMD) during the ionization process in external fields of relativistic intensity \cite{Chowdhury_2001,Dammasch_2001,Yamakawa_2003,Maltsev_2003,Gubbini_2005,DiChiara_2008,Palaniyappan_2008,DiChiara_2010,Ekanayake_2013}. It was clearly shown that the drift of the electron induced by the laser magnetic field  suppresses the usual electron correlation channel -- the recollision, and related phenomena of high-order harmonic generation, above-threshold ionization, and nonsequential double ionization, see e.g. \cite{Dammasch_2001,Kohler_adv,Klaiber_2017}. However, still it is not clarified whether specific electron correlations in the relativistic regime of strong-field ionization, such as shake-up, shake-off processes,  collective tunneling, \textit{etc.} \cite{Joachain_2019}, would exist at suppressed recollisions.

The workhorses of analytical investigations in strong-field physics, the strong-field approximation (SFA) \cite{Keldysh_1965,Faisal_1973,Reiss_1980}, and the quasiclassical imaginary time method (ITM) \cite{Perelomov_1966a,Popov_2004u} have been generalized into the relativistic regime in \cite{Reiss_1990,Reiss_1990b}, and \cite{Popov_1997,Mur_1998,Milosevic_2002r1,Milosevic_2002r2}, respectively. However, in the standard SFA, the influence of the Coulomb field of the atomic core is neglected in the electron continuum dynamics. This approximation is especially unsuitable in the case of HCI.

In the non-relativistic regime the ITM has been improved to treat Coulomb field effects during the ionization  and the well-known quantitatively correct Perelomov-Popov-Terent'ev (PPT) ionization rates have been derived~\cite{Perelomov_1967a}  (in the adiabatic regime also known as Ammosov, Delone, Krainov (ADK) rates \cite{ADK}). The PPT theory uses the quasi-classical wave function for the description of the tunneling part of the electron wave packet through the nonadiabatic barrier formed by the laser and the atomic field. The continuum wave function is matched  to the exact bound state \cite{Perelomov_1966a,Popov_1967}, in this way removing the singularity in the phase of the quasi-classical wave function at the Coulomb center. The PPT theory does not address  Coulomb effects during the photoelectron dynamics in the continuum. The latter is very important for the forming of various features of the photoelectron momentum distribution (PMD) and has been treated within different versions of the Coulomb-corrected SFA (CCSFA). The simplest approach is the Coulomb-Volkov ansatz when the Volkov wave function \cite{Volkov_1935} in the SFA matrix element is replaced by the Coulomb-Volkov wave function \cite{Jain_Tzoar_1978,Yudin_2006}, which incorporates  the asymptotic phase of the exact Coulomb continuum wave function  into the phase of the Volkov state. While the  Coulomb-Volkov approach can be formulated rigorously as an S-matrix expansion \cite{Faisal_2016}, it accounts for the coupling between the Coulomb and the  laser field perturbatively and the approach fails when the electron appears in the continuum after tunneling close to the atomic core~\cite{Smirnova_2006b}.

The extension of the nonrelativistic PPT theory to treat the Coulomb effects in the continuum also employs the electron continuum wave function in the eikonal approximation \cite{Gersten_1975,Krainov_1997,Chirila_2005}. The CCSFA via the eikonal approximation has been rigorously formulated in \cite{Popruzhenko_2008a,Popruzhenko_2008b}, evaluating the eikonal phase of the continuum electron wave function along the exact classical electron trajectories driven by the laser and the Coulomb field. The same approximation has been worked out in \cite{Lai_2015,Maxwell_2017} via the Feynman path integration concept. Higher-order contributions in CCSFA have been discussed in \cite{Klaiber_2017b}, by removing the Coulomb singularity with the use of the saddle-point approximation \cite{Klaiber_2013_I}, rather than with the matching procedure to the bound state.

An innovative way of  matching  the electron eikonal wave function for the continuum to the atomic bound state within the nonrelativistic SFA approach has been advanced in the analytical R-matrix (ARM) theory \cite{Smirnova_2008,Torlina_2012,Torlina_2013,Kaushal_2013}. Here, it has been shown that the rigorous matching procedure is equivalent to a particular (imaginary) shift of the starting point of the complex time-integration in the phase of the eikonal wave function in the SFA amplitude. The ARM theory provides the most efficient version of CCSFA. While the employed eikonal approximation in different versions of CCSFA allows the treatment of rescattering effects, however, restricts the rescattering only to soft ones \cite{Keil_2016}.

The relativistic regime of strong-field ionization can be characterized  by the following parameters. For the sub-barrier dynamics, the parameter $\upsilon\equiv \kappa/c\sim 1$ indicates the relativistic domain, with the atomic momentum $\kappa=\sqrt{2I_p}$, the ionization energy $I_p$, and the speed of light $c$. For the continuum dynamics, the relativistic domain is achieved when the relativistic invariant field parameter $\xi\equiv E_0/(c\omega) \sim 1$, with the laser field amplitude $E_0$, and the frequency $\omega$.  Recollisions in the relativistic regime are suppressed when the Lorentz deflection parameter $\Gamma_R\gtrsim 1$, with  $\Gamma_R\equiv (1/16)\upsilon \xi^3(c^2/\omega)$ \cite{Palaniyappan_2006,Klaiber_2017}. Atomic units are used throughout.

The relativistic domain of strong field ionization is accessible with HCI driven by ultrastrong lasers fields. The ITM including Coulomb corrections during ionization has been extended into the relativistic regime \cite{Popov_1997,Mur_1998,Milosevic_2002r1,Milosevic_2002r2}, allowing to calculate quantitatively relevant ionization rates in the relativistic case. The relativistic version of the plain SFA has been put forward by Howard Reiss in \cite{Reiss_1990, Reiss_1990b}. The CCSFA, based on the relativistic  eikonal-Volkov wave function for the continuum electron \cite{Avetissian_1999}, has been  proposed in \cite{Avetissian_2001}. The calculation of spin-resolved ionization probabilities in the relativistic regime using relativistic CCSFA has been provided in Ref.~\cite{Klaiber_2013_II}, showing the equivalence of the CCSFA to the Coulomb corrected ITM.

We indicate also the significant efforts in the numerical investigations of the relativistic ionization dynamics via the Dirac equation, in particular with HCIs and superstrong laser  fields, carried out in Refs.~\cite{Hu_1999,Casu_2000,Hu_2002,Walser_2002,Mocken_2004b,Mocken_2008,Hetzheim_2009,Selsto_2009,
Vanne_2012,Fillion-Gourdeau_2012,Kjellsson_2017a,Hafizi_2017,Telnov_2020}.

In this paper the ARM theory is extended into the relativistic domain aiming at the application of strong-field ionization of HCIs in ultrastrong laser fields. The ARM theory \cite{Smirnova_2008,Torlina_2012,Torlina_2013,Kaushal_2013} is a version of the eikonal approximation in the description of the Coulomb field of the atomic core for the electron during its continuum dynamics after ionization in a strong laser field. The main advantage of the ARM theory is that the explicit matching procedure of the continuum wave function to the bound state is replaced by the specific shift of the border of the time integration into the complex plane in the eikonal. The  consequence of this procedure is that the singularity of the wave function at the saddle-point of the time integration (corresponding to the center of the Coulomb potential) is eliminated. For the extension of the ARM theory into the relativistic regime we make use of the latter property, namely, derive such a shift of the time integration border in the complex domain which eliminates the singularity  of the phase of the relativistic CCSFA wave function at the time saddle-point.  Finally, we apply the relativistic ARM theory for the investigation of the Coulomb enhancement effect at the cutoff of the direct ionization channel in the relativistic domain with HCIs. This effect is known in the nonrelativistic regime, described in \cite{Keil_2016,He_2018}.

The structure of the paper is the following. In Sec.~\ref{sec:nonrelativistic} we begin with the nonrelativistic regime, elucidating  our approach for the derivation of the ARM theory amplitude, then apply it for the relativistic case in Sec.~\ref{sec:relativistic}. Examples of the application of the derived relativistic ARM-theory are discussed in Sec.~\ref{sec:HECEHIC}, and our conclusions are formulated in Sec.~\ref{sec:conclusion}.

\section{Nonrelativistic theory}\label{sec:nonrelativistic}

In this section we elucidate our approach for the derivation of the ARM-theory amplitude in the nonrelativistic regime. Note that the ionization amplitude in the nonrelativistic ARM-theory has been derived in \cite{Smirnova_2008,Torlina_2012,Torlina_2013,Kaushal_2013} by dividing the interaction region into two sub-regions (inner region and outer region), and by rigorously matching the eikonal wave function for the continuum electron in the outer region to the bound state wave function in the inner-region using a R-matrix approach. Here, we use an operational approach, namely, taking into account that the above mentioned matching procedure of the wave functions is equivalent to a  shift of the time integration border in the complex domain, and we derive such a  shift of the time integration border in the eikonal that eliminates the singularity  of the SFA amplitude. Firstly, we derive with our operational approach the strong-field ionization amplitude in the case of a short-range atomic potential and then discuss the case with the Coulomb field.

We apply SFA for the description of the laser driven ionization process of the atomic bound electron.  The SFA ionization amplitude  of the  electron with the asymptotic outgoing momentum $\mathbf{p}$ is given by \cite{Becker_2002}
\begin{eqnarray}
\label{mmm}
m_\mathbf{p}=-i \int dt \langle \psi_\mathbf{p}(t)|H_i(t)|\phi(t)\rangle,
\end{eqnarray}
where $\phi(t)$ is the  bound state wave function, $\psi_\mathbf{p}(t)$ the electron outgoing continuum state, and  the interaction Hamiltonian in the length gauge
\begin{eqnarray}
H_i(t)=\mathbf{r}\cdot\mathbf{E}(t),
\end{eqnarray}
with the laser electric field $\mathbf{E}=-\partial_t \mathbf{A}$.

\subsection{Short-range potential}

 Let us firstly derive the analytical expression of the ionization amplitude in the leading order $E_0/E_a$ term in the case of a short-range atomic potential, where $E_a=\kappa^3$ is the atomic field. In this case the continuum state in the  laser field in Eq.~(\ref{mmm}) is described by the Volkov-state \cite{Volkov_1935},  $\psi_\mathbf{p}(\mathbf{r},t) \rightarrow \psi^{(0)}_\mathbf{p}(\mathbf{r},t)$:
\begin{eqnarray}
\psi^{(0)}_\mathbf{p}(\mathbf{r},t)=\frac{1}{\sqrt{(2\pi)}^3}\exp\left[i(\mathbf{p}+\mathbf{A}(t))\cdot\mathbf{r}+i\int_t ds\frac{\left (\mathbf{p}+\mathbf{A}(s)\right)^2}{2}\right],\nonumber\\
\end{eqnarray}
with the laser vector potential $\mathbf{A}(t)$, here we consider a linearly polarized laser field. The bound state in the case of the short-range potential is $\phi(\mathbf{r},t)\rightarrow\phi^{(0)}(\mathbf{r},t)$:
\begin{eqnarray}
\phi^{(0)}(\mathbf{r},t)=\sqrt{\frac{\kappa}{2\pi r^2}}\exp\left[-\kappa r+i\frac{\kappa^2}{2}t \right].
\end{eqnarray}
 In this case we straightforwardly arrive at the amplitude $m_\mathbf{p}\rightarrow m^{(0)}_\mathbf{p}$:
\begin{eqnarray}
\label{m000}
m^{(0)}_\mathbf{p}&=&-\frac{i\sqrt{\kappa}}{4\pi^2}\int d^3\mathbf{r}\int dt  \frac{\mathbf{r}\cdot\mathbf{E}(t)}{r}\exp\left[-i\left(\mathbf{p}+\mathbf{A}(t)\right)\cdot\mathbf{r}\right.\nonumber\\
&-&i\left. \int_t ds \frac{ \left(\mathbf{p}+\mathbf{A}(s)\right)^2}{2} +i \frac{\kappa^2}{ 2}t-\kappa r\right].
\end{eqnarray}
In the next step we approximate the $t$-integration via the saddle point approximation (SPA). Here  the solution of the $t$-saddle point equation
\begin{eqnarray}
\left(\mathbf{p}+\mathbf{A}(t)\right)^2+\kappa^2+2\mathbf{r}\cdot\mathbf{E}(t)=0,
\end{eqnarray}
is found perturbatively with respect to the last term which is eqivalent to an expansion in the parameter $E_0/E_a$. It  yields
\begin{eqnarray}
\label{t0}
t_s=\tilde{t}_0+i\frac{\mathbf{r}\cdot\mathbf{E}(\tilde{t}_0)}{|\mathbf{E}(\tilde{t}_0)|\tilde{\kappa}}\approx \tilde{t}_0+i\frac{\mathbf{r}\cdot\mathbf{E}(t_0)}{|\mathbf{E}(t_0)|\kappa},
\end{eqnarray}
where $\tilde{\kappa}=\sqrt{\kappa^2+p_\perp^2}$, and $\tilde{t}_0$ is the common zeroth-order solution \cite{Gribakin_1997} via
\begin{eqnarray}
 \left(\mathbf{p}+\mathbf{A}(\tilde{t}_0)\right)^2+\kappa^2=0.
\end{eqnarray}
Here, we distinguish between $\tilde{t}_0$ and $t_0=\tilde{t}_{0}(\mathbf{p}_{max})$, with $\mathbf{p}_{max}=(-A(t_{max}),0,0)$ the most probable quasi-classical momentum in the nonrelativistic case, corresponding to the ionization at the time $t_{max}$. In the perturbation term in Eq.~(\ref{t0}), we approximate  $\tilde{\kappa}\approx \kappa$ and $\tilde{t}_0\approx t_0$, because otherwise higher order terms with respect to $E_0/E_a$ would be included.
Note that the time dependence of the pre-exponential in Eq.~(\ref{m000}) $\partial_t \ln(E(t))\sim \omega$ is small and can be neglected with an accuracy of $\omega/I_p$ .
With the solution Eq.~(\ref{t0}), the  amplitude in SPA yields
\begin{eqnarray}
\label{mt0}
m^{(0)}_\mathbf{p}&=&-i\int d^3\mathbf{r}\,{\cal M}^{(0)}(\mathbf{r}),\\
{\cal M}^{(0)}(\mathbf{r})&=&\frac{1}{4\pi^2}\sqrt{\frac{2\pi}{|\mathbf{E}(t_0)|}}\frac{ \mathbf{r}\cdot \mathbf{E}(t_0)}{r}\\
&\times&\exp\left[-i\left(\mathbf{p}_{max}+\mathbf{A}(t_0)\right)\cdot\mathbf{r}-\frac{\left(\mathbf{r}\cdot\mathbf{E}(t_0)\right)^2}{2\kappa |\mathbf{E}(t_0)|}-\kappa r\right.\nonumber\\
&-&i \left.\int_{\tilde{t}_0} ds \frac{1}{2}\left(\mathbf{p}+\mathbf{A}(s)\right)^2+i \frac{\kappa^2}{2}\tilde{t}_0\right],\nonumber
\end{eqnarray}
where the terms up to the first order in $E_0/E_a$ in the exponent  are kept, and  with the same accuracy, the pre-factor is estimated at $\mathbf{p}=\mathbf{p}_{max}$.
The remaining $\mathbf{r}$-integral is then calculated analytically:
\begin{eqnarray}
m^{(0)}_\mathbf{p}=\frac{i}{\sqrt{2\pi |\mathbf{E}(t_0)|}}\exp\left[-i\int_{\tilde{t}_0} ds \frac{(\mathbf{p}+\mathbf{A}(s))^2}{2}+i \frac{\kappa^2}{ 2}\tilde{t}_0\right].\nonumber\\
\end{eqnarray}

\subsection{Coulomb potential}

Now with our operational approach we derive the strong-field ionization amplitude in the case of  the atomic Coulomb potential. In this case, the bound state as well as the continuum state in the eikonal approximation obtain exponential  corrections proportional to $\nu=Z/\kappa$, with the charge $Z$ of the atomic core. The bound state in the Coulomb potential in $r\rightarrow \infty$ asymptotics reads
\begin{eqnarray}
\phi (\mathbf{r},t)&=&\phi^{(0)}(\mathbf{r},t)\phi^{(1)}(\mathbf{r}),\\
\phi^{(1)}(\mathbf{r})&=&{\cal C}\exp\left[\nu\ln(\kappa r)\right],\nonumber
\end{eqnarray}
where ${\cal C}$ is the normalization constant (for hydrogen like ions with $Z=\kappa$, it is ${\cal C}=\sqrt{2}$) and the continuum state in the eikonal approximation is
\begin{eqnarray}
\psi (\mathbf{r},t)&=&\psi^{(0)}(\mathbf{r},t)\psi^{(1)}(\mathbf{r}),\\
\psi^{(1)}(\mathbf{r},t)&=&\exp\left[i\nu \int_t ds\frac{\kappa}{|\mathbf{r}+\mathbf{p}(s-t)+\boldsymbol{\alpha}(s)-\boldsymbol{\alpha}(t)|}\right],\nonumber
\end{eqnarray}
where $\boldsymbol{\alpha}(t)=\int  dt \mathbf{A}(t)$. The $t$-dependence in the Coulomb correction (CC) terms is weak and can be neglected with an accuracy of $\omega/I_p$. In this case the CC momentum amplitude of Eq.~(\ref{mt0}) reads,  cf.~\cite{Torlina_2012}:
 \begin{eqnarray}
\label{m111}
m^{(1)}_\mathbf{p}&=&-i \int d^3\mathbf{r}\,{\cal M}^{(0)}(\mathbf{r})\,{\cal C}\exp\left\{\nu\left[\ln(\kappa r)\right.\right.\label{mtsC}\\
&+&i\left.\left.\int_{t_s} ds\frac{\kappa}{|\mathbf{r}+\mathbf{p}(s-t_s)+\boldsymbol{\alpha}(s)-\boldsymbol{\alpha}(t_s)|}\right]\right\}.\nonumber
\end{eqnarray}

\begin{table}[t]
\begin{center}
\begin{tabular}{c| c |c|c}
&quantity & non-relativsitic estimate  & relativistic estimate \\
\hline
&$r_E$&$ \sqrt{E_a/E_0}/\kappa $ &$  g(I_p/c^2)\sqrt{E_a/E_0}/\kappa$\\[8pt]
&$r_k$ &$0$ &0\\[8pt]
&$r_B$&$0$ &0\\[8pt]
&$p_{k0}$&0 & $c(\lambda^2-1)/(2\lambda)$\\[8pt]
&$p_{E0}$&0 &0\\[8pt]
&$p_{B0}$ &0&0\\[8pt]
&$\Delta  p_k$&$\sqrt{E_0/E_a}\,\kappa$ &$g(I_p/c^2 )\sqrt{E_0/E_a} \, \kappa$\\[8pt]
&$\Delta p_E$&$\sqrt{E_0/E_a} \, E_0/\omega$ &$g(I_p/c^2)\sqrt{E_0/E_a}  \,E_0/\omega$\\[8pt]
&$\Delta  p_B$ &$\sqrt{E_0/E_a} \,\kappa$ &$g(I_p/c^2)\sqrt{E_0/E_a}\, \kappa $\\[8pt]
\end{tabular}
\label{table1}
\end{center}
\caption{Estimation of the variables and parameters via SFA for the approximate calculation of the integral in Eq.~(\ref{delta}), where $\mathbf{r}=(r_k,r_E,r_B)$, $\mathbf{p}=(p_k,p_E,p_B)$, the components of the vectors are defined along the laser propagation direction, the laser electric field, and along the the laser magnetic field,  the function $g(I_p/c^2)$ depends on $I_p/c^2$. }
 \end{table}

The $s$-integral in the phase of Eq.~(\ref{mtsC}) is diverging at the low limit $t=t_s$ at the Coulomb center $\mathbf{r}=0$. However, the diverging term can be canceled with the bound CC term $\phi^{(1)}(\mathbf{r})$, when using appropriate approximations. We separate the diverging term in  integral of Eq.~(\ref{mtsC}) $\int_{t_s} =\int_{t_s}^{t_0-i\delta}+\int_{t_0-i\delta}$, and show that with appropriate choice of the parameter $\delta$, the  diverging term $\int_{t_s}^{t_0-i\delta}$ will be canceled by the bound CC term when using the following approximations.
Firstly, in the analytical calculation of CC in the continuum state, we neglect  all higher order corrections with respect to $\sqrt{E_0/E_a}$,  in the spirit of ARM \cite{Torlina_2012}. To apply the given approximation, we estimate the variables in the integrand of Eq.~(\ref{mtsC}), with the result summarized in Table~I. 
We refer to \cite{Klaiber_2013_I} [below Eq.~(21)] for the estimation $r_E=\sqrt{E_a/E_0}/\kappa $, which is the scaling of the coordinate saddle point of the integrand in Eq.(\ref{mt0}), i.e., the point, where the ionizing trajectory starts;  $r_k\approx r_k(t_0)=0$ is approximated, assuming that the most probable trajectory has zero impact parameter at $t_0$ near the core. Further, we express $p_{k,E,B}=p_{k,E,B\,0}+\Delta p_{k,E,B}$, with the most probable value of the momentum $p_{k,E,B\,0}$ [in the nonrelativistic theory $p_{k,E,B\,0}=0$, and in the relativistic one $p_{k0}= c(\lambda^2-1)/(2\lambda)$, $p_{E,B\,0}=0$, see Eq.~(\ref{pko}) below], and the new variables $\Delta p_{k,E,B}$ corresponding to the momentum width of the tunneling wave packet. The latter are estimated as $\Delta p_{E}\sim\sqrt{E_0/E_a} E_0/\omega$, $\Delta p_{k}=\Delta p_{B}\sim\sqrt{E_0/E_a} \kappa$  according to the PPT theory \cite{Popov_2004u}. In the relativistic estimations an additional factor $g(I_p/c^2$ depending on $I_p/c^2$ arises. Here, $\mathbf{r}=(r_k,r_E,r_B)$, $\mathbf{p}=(p_k,p_E,p_B)$, and the components of the vectors are defined along the laser propagation direction $r_k\equiv \mathbf{r}\cdot\mathbf{k}/k$, $p_k\equiv \mathbf{p}\cdot\mathbf{k}/k$, along the laser electric field $r_E\equiv -\mathbf{r}\cdot\mathbf{E}/E_0$, $p_E\equiv -\mathbf{p}\cdot\mathbf{E}/E_0$, and along the laser magnetic field $r_B\equiv -\mathbf{r}\cdot\mathbf{B}/B_0$, $p_B\equiv -\mathbf{p}\cdot\mathbf{B}/B_0$.
We introduce dimensionless variables $R_E,\,P_k,\,P_E,\,P_B$, dividing the given variable over its estimated value in Table~I: $R_E\equiv r_E\kappa \sqrt{E_0/E_a}$,$P_k\equiv \Delta p_k/(\sqrt{E_0/E_a}\kappa),\, P_E\equiv \Delta p_E/(\sqrt{E_0/E_a}E_0/\omega),\, P_B\equiv \Delta p_B/(\sqrt{E_0/E_a}\kappa)$. 
Further, we apply a variable transformation $s=t_s+\sigma(t_0-i \delta-t_s)$
and expand the integrand up to leading order in $E_0/E_a$ in a quasistatic approximation arriving at:
\begin{eqnarray}
&&\exp\left[i\nu\int^{t_0-i\delta}_{t_s} ds\frac{\kappa}{|\mathbf{r}+\mathbf{p}(s-t_s)+\boldsymbol{\alpha}(s)-\boldsymbol{\alpha}(t_s)|}\right]\nonumber\\
\approx&&\exp\left[\nu \int^1_0 d\sigma \left(\frac{\sqrt{E_0/E_a} \left(2\delta\kappa^2-(\sigma -1)^2 R_E^2\right)}{2 (\sigma -1)^2 R_E}+\frac{1}{\sigma -1}\right)\right]\nonumber\\
&&+{\cal O}(\sqrt{E_0/E_a})\nonumber\\
\approx &&\left[ \delta \kappa^2 \left(1/R_E-\sqrt{E_0/E_a}\right)\sqrt{E_0/E_a}\right]^\nu +{\cal O}(\sqrt{E_0/E_a})\nonumber\\
\approx&&\left(\delta\kappa/r_E\right)^{\nu}+{\cal O}(\sqrt{E_0/E_a}).
\label{delta}
\end{eqnarray}
The singular term at $r\rightarrow 0$ in Eq.~(\ref{delta})  will be canceled with the similar term in the Coulomb correction of the bound state, see Eq.~(\ref{m111}), if we  choose $\delta=1/\kappa^2$. 
Secondly, we approximate
\begin{eqnarray}
&&|\mathbf{r}+\mathbf{p}(s-t_s)+\boldsymbol{\alpha}(s)-\boldsymbol{\alpha}(t_s)|\approx|\mathbf{p}_{max}(s-t_0)+\boldsymbol{\alpha}(s)-\boldsymbol{\alpha}(t_0)\nonumber\\
&+&{\cal O}(\sqrt{E_0/E_a})|,\nonumber\\
\end{eqnarray}
which  again follows from the scaling laws given in Table~I.

Consequently, we obtain that the correction terms are approximately independent of the coordinates and arrive at the momentum amplitude:
 \begin{eqnarray}
m^{(1)}_\mathbf{p}&=&-i {\cal C}\int d^3\mathbf{r} {\cal M}^{(0)}(\mathbf{r})\\
&\times &\exp\left[ i \int_{t_0-\frac{i}{\kappa^2}} \frac{Z\,\,ds}{|\mathbf{p}_{max}(s-t_0)+\boldsymbol{\alpha}(s)-\boldsymbol{\alpha}(t_0)|}\right].\nonumber
\label{mtC}
\end{eqnarray}
The latter after the final coordinate integration yields
\begin{eqnarray}
m^{(1)}_\mathbf{p}={\cal C}m^{(0)}_\mathbf{p}\exp\left[i\int_{t_0-\frac{i}{\kappa^2}}  \frac{Z \,\,ds}{|\mathbf{p}_{max}(s-t_0)+\boldsymbol{\alpha}(s)-\boldsymbol{\alpha}(t_0)|}\right].\nonumber\\
\end{eqnarray}

\section{Relativistic theory}\label{sec:relativistic}

In the relativistic regime we employ SFA based on the Dirac equation \cite{Klaiber_2013_II}. The ionization SFA amplitude is again formally given by Eq.~(\ref{mmm}), where the interaction Hamiltonian in the G\"oppert-Mayer gauge within the dressed partition \cite{Klaiber_2013_II} reads:
\begin{eqnarray}
H_i(\mathbf{r},t)&=&\mathbf{r}\cdot\mathbf{E}(\eta),\\
H_0&=&H_a-\mathbf{r}\cdot\mathbf{E}(\eta)\alpha_k
\end{eqnarray}
where $\alpha_k\equiv \bm{\alpha}\cdot \hat{\mathbf{k}}$, $\bm{\alpha}$ are Dirac matrices, $\hat{\mathbf{k}}$  is the unit vector along the laser propagation direction,  and $\eta=t-\hat{\mathbf{k}}\cdot\mathbf{r}/c$. The spin quantization axis is chosen  along the laser magnetic field. We employ the four vector potential of the laser field in the G\"oppert-Mayer gauge: $A^\mu=-\left(\hat{\mathbf{k}}(\mathbf{r}\cdot\mathbf{E}(\eta),\mathbf{r}\cdot\mathbf{E}(\eta)\right)$.
In \cite{Klaiber_2013_II} we have shown that the relativistic SFA provides more close expressions to the relativistic PPT theory \cite{Milosevic_2002r2} for the total ionization rate if the dressed partition is applied. In the dressed partition, the unperturbed bound state is corrected \cite{Klaiber_2013_II} by a factor
 \begin{eqnarray}
{\cal S}=\exp\left(i\frac{A}{2c-I_p/c}\right).
\end{eqnarray}
\\

\subsection{Short-range potential}

In the case of a short-range potential the outgoing state is the relativistic Volkov state $\psi_\mathbf{p}(\mathbf{r},t) \rightarrow \psi^{(0)}_\mathbf{p}(\mathbf{r},t)$:
 \begin{eqnarray}
&&\psi^{(0)}_\mathbf{p}(\mathbf{r},t)=\left(1+\frac{(1+\alpha_k){\boldsymbol \alpha}\cdot\mathbf{A}(\eta)}{2c\tilde{\Lambda}}\right)\frac{cu_f}{\sqrt{(2\pi)^{3}\tilde{\varepsilon}}}\\
&\times&\exp\left[i(\mathbf{p}+\mathbf{A}(\eta))\cdot\mathbf{r}-i \tilde{\varepsilon} t+i\int_\eta ds\left(\frac{\mathbf{p}\cdot\mathbf{A}(s)+A(s)^2/2}{\tilde{\Lambda}}\right)\right],\nonumber
\end{eqnarray}
with the asymptotic energy $\tilde{\varepsilon}=\sqrt{c^4+c^2{\mathbf p}^2}$, the constant of motion $\tilde{\Lambda}=\tilde{\varepsilon}/c^2-p_k/c$, and the bispinor
\begin{eqnarray}
u_f=\left(\sqrt{\frac{c^2+\tilde{\varepsilon}}{2c^2}}\chi_f,\frac{{\boldsymbol\sigma}\cdot\mathbf{p}}{\sqrt{2(c^2+\tilde{\varepsilon})}}\chi_f\right)^T,
\end{eqnarray}
where $\chi_+=(1,0)^T$ and $\chi_-=(0,1)^T$. The bound state of the short-range potential is
\begin{eqnarray}
\phi^{(0)}(\mathbf{r},t)=\sqrt{\frac{\kappa}{2\pi r^2}}\exp\left[-\kappa r+i (I_p-c^2) t \right]v_i,
\end{eqnarray}
with the atomic momentum $\kappa=\sqrt{2I_p\left(1-\frac{I_p}{2c^2}\right)}$, and the  bispinor
\begin{eqnarray}
v_i=\left(\chi_i,i\frac{c(\kappa r+1){\boldsymbol\sigma}\cdot\mathbf{r}}{(2c^2-I_p)r^2}\chi_i\right)^T.
\end{eqnarray}

We consider two cases, firstly, when there is no spin flip: $\chi_f=\chi_i=\chi_+$, and secondly, when there is  a spin flip during ionization, i.e. $\chi_f=\chi_-$ and $\chi_i=\chi_+$. In the first case we have
\begin{eqnarray}
m^{(0)}_{\mathbf{p}+}&=& -\frac{i\left(2I_p\right)^{1/4}}{4\pi^2}\int d^3\mathbf{r}d\eta \,{\cal S}(\eta)\,{\cal P}_+\frac{\mathbf{r}\cdot\mathbf{E}(\eta)}{r}\\
&\times &\exp\left\{-i\left[\mathbf{p}+\mathbf{A}(\eta)+\frac{c^2-I_p-\varepsilon}{c}\hat{\mathbf{k}}\right]\cdot\mathbf{r}\right.\nonumber\\
&&-i\left.\int_\eta ds\left[\tilde{\varepsilon}+\frac{\mathbf{p}\cdot\mathbf{A}(s)+A(s)^2/2}{\tilde{\Lambda}}\right]+i (I_p-c^2)\eta-\kappa r\right\}\nonumber
\end{eqnarray}
with
\begin{widetext}
\begin{eqnarray}
{\cal P}_+&=&\frac{2 c \Lambda  \left(-c \sqrt{I_p \left(2 c^2-I_p\right)} p_{k0}-c^2 \left(-2 \varepsilon +I_p\right)+2 c^4-I_p \varepsilon \right)+ \sqrt{
2\Lambda(\varepsilon-c^2 + I_p) } \left(\sqrt{-I_p \left(-2 c^2+I_p\right)}+2
   c^2-I_p\right) \left(\varepsilon+c \left(c+p_{k0}\right)\right)}{2 c^{3/2} \Lambda  \left(4 c^2-2 I_p\right){}^{3/4} \sqrt{\varepsilon  \left(c^2+\varepsilon \right)}}\nonumber\\
&&+{\cal O}(\sqrt{E_0/E_a}),
\end{eqnarray}
\end{widetext}
where the amplitude is evaluated at the most probable quasiclassical momentum $\mathbf{p}_{max}=(0,0,p_{k0})$, with 
\begin{eqnarray}
p_{k0}=c(\lambda^2-1)/(2\lambda),\label{pko}
\end{eqnarray}
$\lambda=(\sqrt{\epsilon^2+8}-\epsilon)/2$, $\epsilon=1-I_p/c^2$ and the notations $\Lambda\equiv \tilde{\Lambda}(\mathbf{p}_{max})$ and $\varepsilon \equiv \tilde{\varepsilon}(\mathbf{p}_{max})$ .
In the second case the amplitude reads:
\begin{eqnarray}
m^{(0)}_{\mathbf{p}-}&=& -\frac{i\left(2I_p\right)^{1/4}}{4\pi^2}\int d^3\mathbf{r}d\eta \,{\cal S}(\eta)\,{\cal P}_- \frac{\mathbf{r}\cdot\mathbf{E}(\eta)}{r}\\
&\times & \exp\left\{-i\left[\mathbf{p}+\mathbf{A}(\eta)+\frac{c^2-I_p-\varepsilon}{c}\hat{\mathbf{k}}\right]\cdot\mathbf{r}\nonumber \right.\\
&-& \left.i\int_\eta ds\left[\tilde{\varepsilon}+\frac{\mathbf{p}\cdot\mathbf{A}(s)+A(s)^2/2}{\tilde{\Lambda}}\right]+i (I_p-c^2)\eta-\kappa r\right\},\nonumber
\end{eqnarray}
with ${\cal P}_-=0+{\cal O}(\sqrt{E_0/E_a})$. We have expanded the expressions in the parameter $E_0/E_a$ with the relativstic atomic field $E_a=\kappa^3$ and $\kappa$ along the lines of SFA. From the expansion it follows that in leading order in this parameter no spin flip occurs and, consequently, we focus only on the spin flip free process.

In the next step we approximate the $\eta$-integration via SPA. Here,  the $\eta$-saddle point equation is
\begin{eqnarray}
\varepsilon+\frac{\mathbf{p}\cdot\mathbf{A}(\eta)+A(\eta)^2/2}{\Lambda}+I_p-c^2+\mathbf{r}\cdot\mathbf{E}(\eta)=0,
\end{eqnarray}
which is solved perturbatively with respect to the last term, yielding the solution
\begin{eqnarray}
 \eta_s=\tilde{\eta}_0+\frac{\mathbf{r}\cdot\mathbf{E}(\eta_0)\Lambda}{ [(\mathbf{p}_{max}+\mathbf{A}(\eta_0))\cdot\mathbf{E}(\eta_0)]},
\end{eqnarray}
with the  zeroth order solution $ \tilde{\eta}_0$ and $\eta_0=\tilde{\eta}_0(\mathbf{p}_{max})$ \cite{Klaiber_2013_II}.
The time dependence of the pre-exponential  $\partial_\eta \ln(E(\eta))\sim \omega$ is small and can be neglected.
After the $\eta$-SPA, the amplitude is
\begin{widetext}
\begin{eqnarray}\label{mt}
 m^{(0)}_{\mathbf{p}+}&=&-i\int d^3\mathbf{r}{\cal M}^{(0)}(\mathbf{r})\\
 {\cal M}^{(0)}(\mathbf{r})&=&\frac{\left(2I_p\right)^{1/4}}{4\pi^2}{\cal S}(\eta_0){\cal P}_+\sqrt{\frac{-2i\pi \Lambda}{\left(\mathbf{p}_{max}+\mathbf{A}(\eta_0)\right)\cdot\mathbf{E}(\eta_0)}}\frac{ \mathbf{r}\cdot \mathbf{E}(\eta_0)}{r}
\exp\Big\{-i\left[\mathbf{p}_{max}+\mathbf{A}(\eta_0)+\frac{\varepsilon-c^2+I_p}{ c}\hat{\mathbf{k}}\right]\cdot\mathbf{r}\nonumber\\
&&+\frac{i\left(\mathbf{r}\cdot\mathbf{E}(\eta_0)\right)^2\Lambda}{
2(\mathbf{p}_{max}+\mathbf{A}(\eta_0))\cdot\mathbf{E}(\eta_0)}-\kappa r
-i \int_{\tilde{\eta}_0} ds\left(\tilde{\varepsilon}+\frac{\mathbf{p}\cdot\mathbf{A}(s)+A(s)^2/2}{\tilde{\Lambda}}\right)+i (I_p-c^2)\tilde{\eta}_0\Big\},
\end{eqnarray}
\end{widetext}
where terms up to the next to leading  order in $E_0/E_a$ in the exponent are kept.
The remaining $\mathbf{r}$-integral is then calculated analytically in leading order in $E_0/E_a$
\begin{eqnarray}
m^{(0)}_\mathbf{p}&=&\frac{i}{\sqrt{2\pi |\mathbf{E}(\eta_0)|}}\,{\cal S}\,{\cal P}_+Q\\
&\times&\exp\left\{-i \int_{\tilde{\eta}_0} ds\left[\tilde{\varepsilon}+\frac{\mathbf{p}\cdot\mathbf{A}(s)+A(s)^2/2}{\tilde{\Lambda}}\right]+i (I_p-c^2)\tilde{\eta}_0\right\}.\nonumber
\end{eqnarray}
with  the pre-factor
\begin{eqnarray}
Q=\sqrt{\frac{\varepsilon-c^2 + I_p}{I_p}}
\end{eqnarray}
also evaluated at the most probable quasi-classical momentum.

\subsection{Coulomb potential}

In the relativistic case the  corrections to the wave functions of the order of $Z/\kappa$ due to the Coulomb potential are the following, for the bound state
\begin{eqnarray}
    \phi (\mathbf{r},\eta)&=&  \phi^{(0)}(\mathbf{r},\eta)\phi^{(1)}(\mathbf{r},\eta) \\
   \phi^{(1)}(\mathbf{r})&=&  {\cal C} \exp\left[\nu\ln(\kappa r)\right],\\
   \nu &=&\frac{(c^2 - I_p) Z}{c^2 \sqrt{I_p (2 - I_p/c^2)}} \label{nunu}
\end{eqnarray}
with the normalization constant  ${\cal C}=2^{\nu -\frac{1}{2}} \sqrt{\frac{\nu +1}{\Gamma (2 \nu +1)}}$ for hydrogen-like systems with $Z=\kappa$, and the continuum state
\begin{eqnarray}
\label{rel_s}
    \psi (\mathbf{r},\eta)&=&  \psi^{(0)}(\mathbf{r},\eta)\psi^{(1)}(\mathbf{r},\eta) \\
   \psi^{(1)}(\mathbf{r},\eta)&=&\exp\left\{i\int_\eta ds\frac{Z\tilde{\varepsilon}(s)}{\tilde{\Lambda} c^2}\right.\\
&\times&\left.\frac{1}{|\mathbf{r}+(\mathbf{p}(s-\eta)+\boldsymbol{\alpha}(s)-\boldsymbol{\alpha}(\eta))/\tilde{\Lambda}+\mathbf{r}_k(s,\eta)|}\right\}\nonumber
\end{eqnarray}
with $\mathbf{r}_k(s,\eta)= \hat{\mathbf{k}}[\mathbf{p}\cdot({\boldsymbol \alpha}(s)-{\boldsymbol \alpha}(\eta))+\beta(s)- \beta(\eta)]/(c\tilde{\Lambda}^2)$, $\beta=\int ds \mathbf{A}^2/2$ and $\tilde{\varepsilon}(\eta)=\tilde{\varepsilon}+(\mathbf{p}\cdot\mathbf{A}(\eta)+A(\eta)^2/2)/\tilde{\Lambda}$.

The singularity in the $s$-integral in the phase of Eq.~(\ref{rel_s}) is removed using the same procedure as in the nonrelativistic case.  The first term in the integral $\int_{\eta_s} =\int_{\eta_s}^{\eta_0-i\delta}+\int_{\eta_0-i\delta}$ is divergent, which is canceled with the bound CC term $\phi^{(1)}(\mathbf{r})$, when using  an appropriate value for the parameter $\delta$.

Taking into account that the $\eta$-dependence in the CC terms is weak, the momentum amplitude for the Coulomb potential is approximated by Eq.~(\ref{mt}) including  extra CC terms:
\begin{eqnarray}
m^{(1)}_{\mathbf{p}+}&=&-i {\cal C}\int d^3\mathbf{r}{\cal M}^{(0)}(\mathbf{r})\exp\left\{\nu\ln(\kappa  r))\right.\\
&+&\left. i\int_{\eta_s}ds\frac{Z\tilde{\varepsilon}(s)}{\tilde{\Lambda} c^2}\frac{1}{|\mathbf{r}+(\mathbf{p}(s-\eta_s)+\boldsymbol{\alpha}(s)-\frac{{\boldsymbol \alpha}(\eta_s)}{\tilde{\Lambda}}+\mathbf{r}_k(s,\eta_s)|}\right\}.\nonumber
\label{mtC}
\end{eqnarray}
The choice of $\delta$ for the singularity removal is possible when the following approximations are applied. We integrate the Coulomb correction in the continuum state analytically around $\eta_s$:
\begin{eqnarray}
&&\exp\left[i\int^{\eta_0-i\delta}_{\eta_s}\frac{Z\tilde{\varepsilon}(s)ds}{\tilde{\Lambda} c^2|\mathbf{r}+(\mathbf{p}(s-\eta_s)+\boldsymbol{\alpha}(s)-\boldsymbol{\alpha}(\eta_s))/\tilde{\Lambda}+\mathbf{r}_k(s,\eta_s)|}\right]\nonumber\\
\approx&&\left(\frac{\sqrt{3}\lambda\kappa\delta}{\sqrt{(4-\lambda^2)}r_E}\right)^\nu,
\end{eqnarray}
where the same method as in the derivation of Eq.~(\ref{delta}) is applied, i.e. the same scaled variables are introduced $R_E$, $P_k$, and $P_E$ $P_B$, the variable transformation $s=\eta_s+\sigma(\eta_0-i \delta-\eta_s)$ is used, and then, before analytical integration the integrand is expanded in $\sqrt{E_0/E_a}$ in a quasistatic approximation, using estimations of Table~I. 
Further, the atomic correction term is in leading order in $E_0/E_a$
\begin{eqnarray}
(\kappa r)^\nu=\left(\kappa\frac{\sqrt{4-\lambda^2}r_E}{\sqrt{3}}\right)^\nu .
\end{eqnarray}
With the choice $\delta=1/(\lambda\kappa^2)=\Lambda/\kappa^2\approx(1-\kappa^2/(6c^2))/\kappa^2$ we cancel  the singular term of the Coulomb correction to the bound state. 
Then, we approximate in leading order in $E_0/E_a$:
\begin{eqnarray}
&&|\mathbf{r}+(\mathbf{p}(s-\eta_s)+\boldsymbol{\alpha}(s)-\boldsymbol{\alpha}(\eta_s))/\tilde{\Lambda}+\mathbf{r}_k(s,\eta_s)|\\
&\approx&|(\mathbf{p}_{max}(s-\eta_0)+\boldsymbol{\alpha}(s)-{\boldsymbol \alpha}(\eta_0))/\Lambda+\mathbf{r}_k(s,\eta_0)+{\cal O}(\sqrt{E_0/E_a})|\nonumber.
\end{eqnarray}
Thus, we conclude that the correction terms are approximately independent of the coordinates, and arrive at the momentum amplitude:
\begin{eqnarray}
m^{(1)}_\mathbf{p}&=&-i{\cal C}\int d^3\mathbf{r} {\cal M}^{(0)}(\mathbf{r}) \\
&\times&\exp\left\{ i \int_{\eta_0-i\Lambda/\kappa^2} \frac{Z\varepsilon(s)/ (c^2\Lambda)\,\, ds}{|\mathbf{p}_{max}(s-\eta_0)+{\boldsymbol \alpha}(s)-{\boldsymbol \alpha}(\eta_0)+\mathbf{r}_k(s,\eta_0)|}\right\}\nonumber
\label{mtC}
\end{eqnarray}
after the final coordinate integration this yields
\begin{eqnarray}
\label{m1R}
m^{(1)}_\mathbf{p}&=&{\cal C}m^{(0)}_\mathbf{p}\\
&\times&\exp\left\{ i \int_{\eta_0-i\Lambda/\kappa^2} \frac{Z\varepsilon(s)/ (c^2\Lambda)\,\, ds}{|\mathbf{p}_{max}(s-\eta_0)+{\boldsymbol \alpha}(s)-{\boldsymbol \alpha}(\eta_0)+\mathbf{r}_k(s,\eta_0)|}\right\}\nonumber
\end{eqnarray}
The equation (\ref{m1R}) is the main result of the paper, providing the Coulomb corrected strong-field ionization amplitude for the relativistic regime using the ARM approach.

\section{ Comparison with the relativistic CCSFA }

\begin{figure}
  \begin{center}
\includegraphics[width=0.5\textwidth]{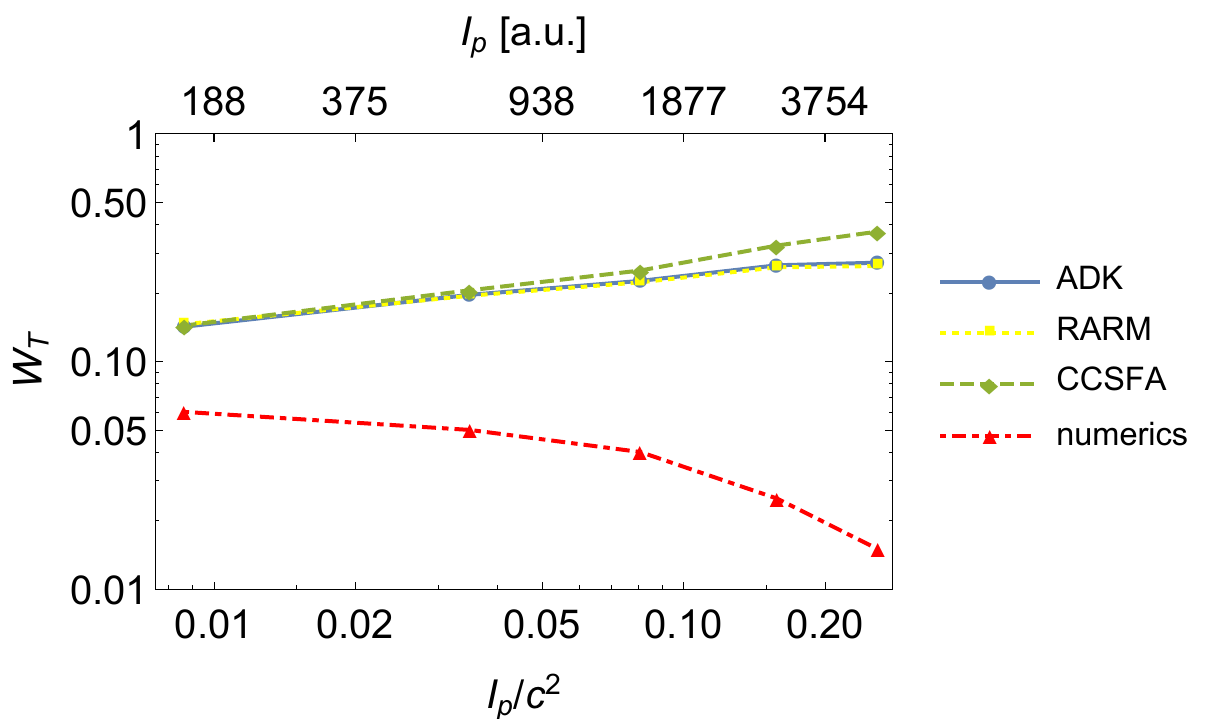}
 \caption{ Comparison of the theoretical data for the total probability $W_T$ per the laser period with the result of the numerical calculation \cite{Hafizi_2017}:  via RARM (yellow short-dashed line with squares) with CCSFA of \cite{Klaiber_2013_I}  (green long-dashed line with diamonds) using dressed partition, ADK (blue solid line) theories at  $Z/\kappa=1$, and  (red dash-dotted line with triangles) the numerical calculations via the Klein-Gordon equation \cite{Hafizi_2017}. Numerical calculations in \cite{Hafizi_2017} have been carried out for the ionization energies  $I_p/c^2 = 0.00866,\, 0.0351, \,0.0809,\,0.158, \,0.259$, using $E_0/E_a\approx 1/16$.}
\label{fig1}
\end{center}
\end{figure}

The comparison of the total ionization probability in relativistic ARM (RARM) with CCSFA of \cite{Klaiber_2013_II} and PPT  theories is provided in Fig.~\ref{fig1}. The RARM probability coincides with the PPT theory, whereas the CCSFA overestimates slightly the PPT theory. Here the rate is calculated in leading order in $E_0/E_a$ and the final momentum integration is accomplished via SPA at the most probable momentum after a transformation from the asymptotic momentum $\mathbf{p}$ to the tunnel exit distribution $(\eta_e, p_{e,B}, p_{e,k})$ with $p_E+A(\eta_e)=0, p_B=p_{e,B}$ and $p_{k}=p_{k,e}+A(\eta_e)^2/2/c/\tilde{\Lambda}$.

\begin{figure}[b]
  \begin{center}
\includegraphics[width=0.45\textwidth]{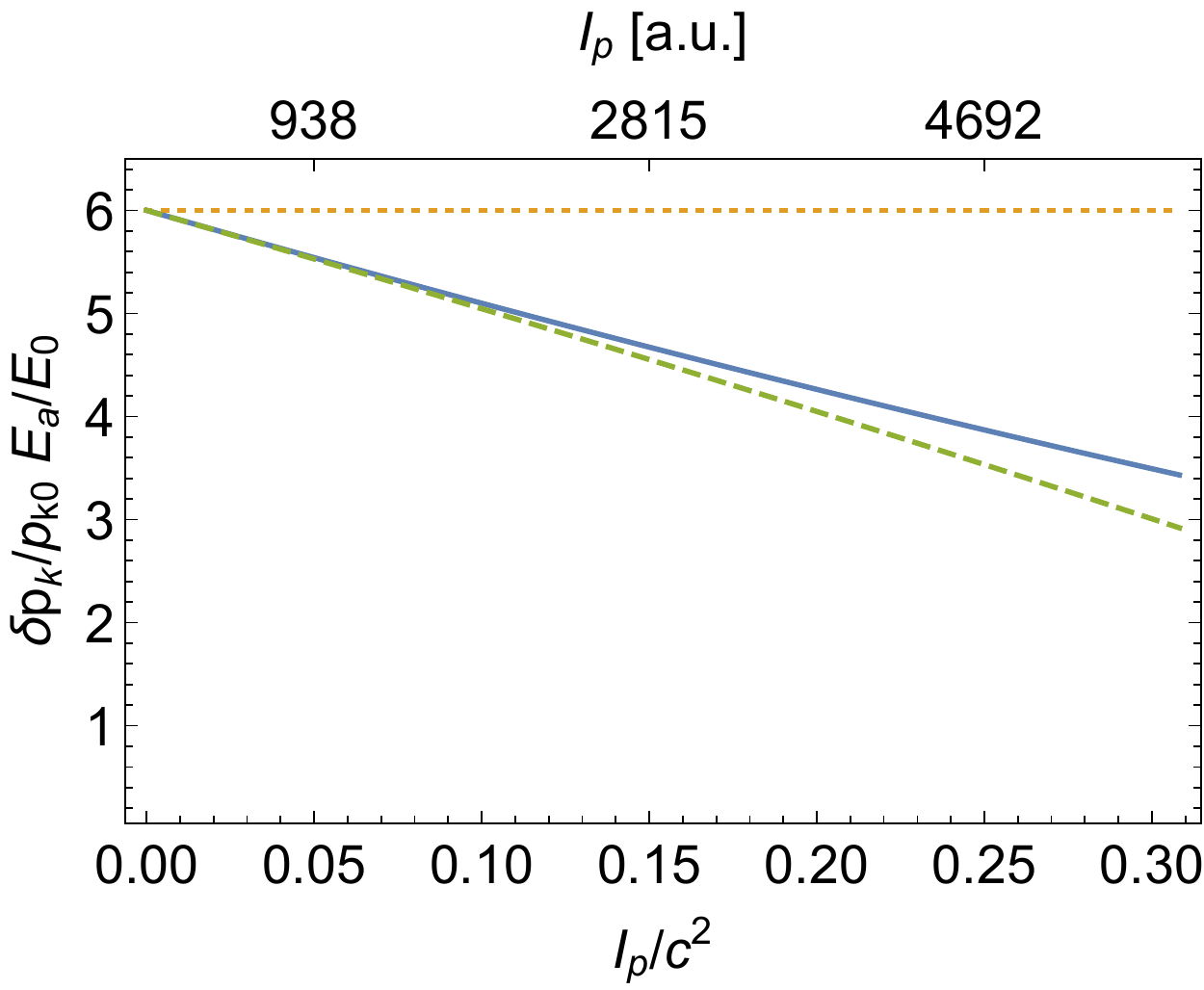}
 \caption{The nondipole shift of the peak of the longitudinal momentum due to the sub-barrier CC for hydrogenlike highly charged ions: via RARM (blue solid) via Eq.~(\ref{pkexact}), via nondipole CCSFA \cite{He_2022} (orange short-dashed), and via the approximate Eq.~(\ref{pk000}) with the leading correction $\sim I_p/c^2$ (green long-dashed) }.
\label{fig2}
\end{center}
\end{figure}

In  Fig.~\ref{fig1}, the comparison of the results for the total probability per laser cycle via RARM, CCSFA, and PPT with the numerical calculation of Hafizi et al. using the Klein-Gordon equation \cite{Hafizi_2017}, is shown. While the theoretical results almost coincide with each other, there is a significant deviation from the numerical calculation especially at high values of $I_p/c^2$.
There are two reasons for the deviation of the analytical quasiclassical theories, CCSFA and RARM, with respect to the numerical result. 
We compare the total ionization rate via RARM based on the Dirac equation with the numerical solution of the Klein-Gordon equation, intuitively assuming that for the total ionization rate spin effects would not matter much. However, this assumption is valid only for $I_p/c^2\ll 1$. The results of Ref.~\cite{Popov_2004u,Klaiber_2014} show that at large $I_p/c^2\sim 1$ spin asymmetry arises in the ionization (difference in ionization probability of different spin states) which will lead to a modification of the spin averaged probability. However, this effect is of the order of at most 1\% even for hydrogenlike uranium and cannot account for the large discrepancy apparent in Fig.~\ref{fig2}.  The main source of the deviation (by a factor of  $\sim 20$) possibly comes from the Stark-shift and polarization of the atomic state in strong fields near the threshold of the over-the-barrier ionization. These corrections are especially relevant in the near-threshold regime of tunneling ionization at $E_0/E_a\sim 1/10$, which is the case in the numerical data of Hafizi ($E_0/E_a\sim 1/16$). 
We assume that this deviation could be corrected, at least partly, via the next order quasiclassical CCs to the eikonal approximation. As is shown in Ref.~\cite{Klaiber_2017b} with 1D CCSFA for the nonrelativistical theory [see Eq.~(47) in this reference], this kind of corrections lead to a decrease of the tunneling ionization probability. The high-order quasiclassical CCs within ARM is generally possible, but it would require the change of the matching procedure to the bound state and, consequently, the change of the complex shift of the time integration.

From a technical point of view RARM has a clear advantage with respect to relativistic eikonal CCSFA of \cite{Klaiber_2013_II,Klaiber_2022} when applying SPA. While in RARM the ionization amplitude is found via a one-dimensional $\eta$-SPA, in CCSFA at least the two-dimensional (in the case of  linear polarization), or four-dimensional (in the case of ellipictical polarization)  SPA for $(\mathbf{r},\eta)$ integrations are required.

Disadvantage of  RARM is that it includes accurately the  CC near the tunnel exit, but overestimates those due to rescatterings. 
To account for CC at hard recollisions, the generalized eikonal approximation (GEA) has been developed on the basis of CCSFA. Generalization to elliptical polarization in both cases (RARM/CCSFA) is possible.

\begin{figure*}
\begin{center}
\includegraphics[width=0.32\textwidth]{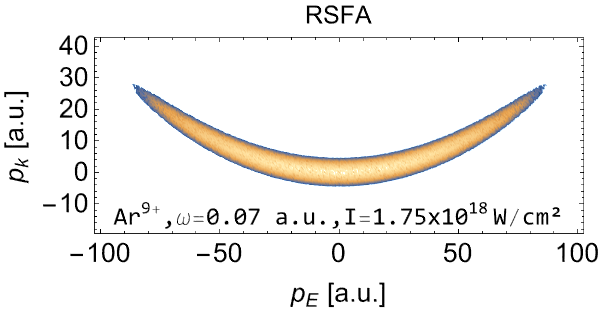}
\includegraphics[width=0.32\textwidth]{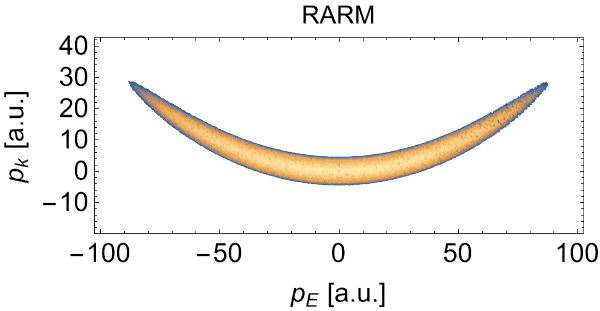}
\includegraphics[width=0.32\textwidth]{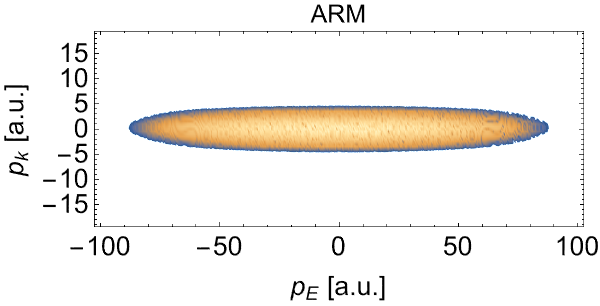}
\includegraphics[width=0.32\textwidth]{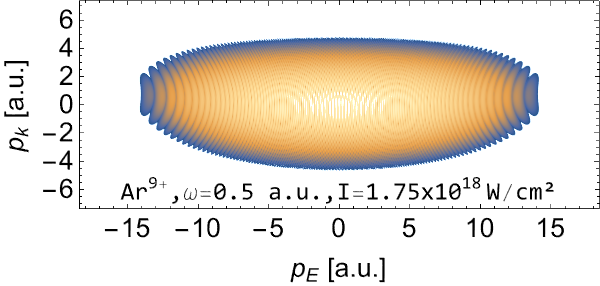}
\includegraphics[width=0.32\textwidth]{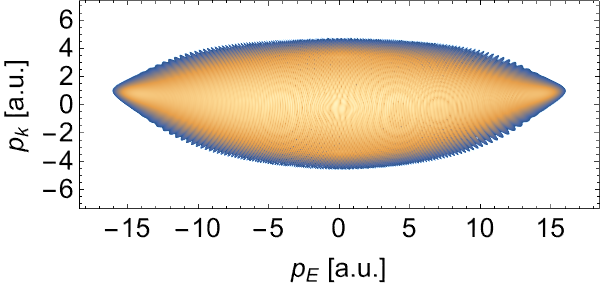}
\includegraphics[width=0.32\textwidth]{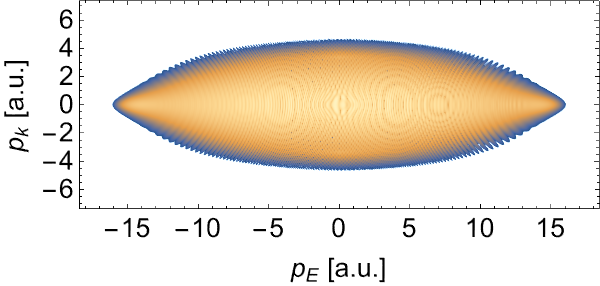}
\includegraphics[width=0.32\textwidth]{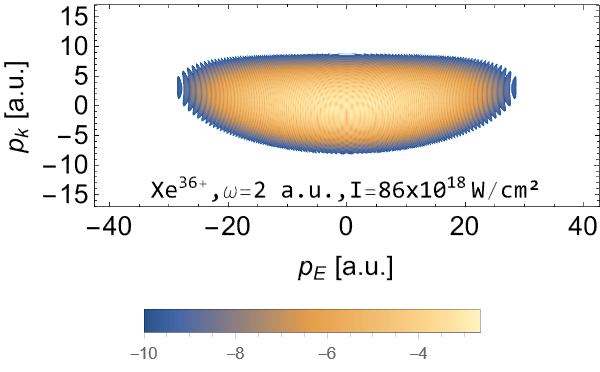}
\includegraphics[width=0.32\textwidth]{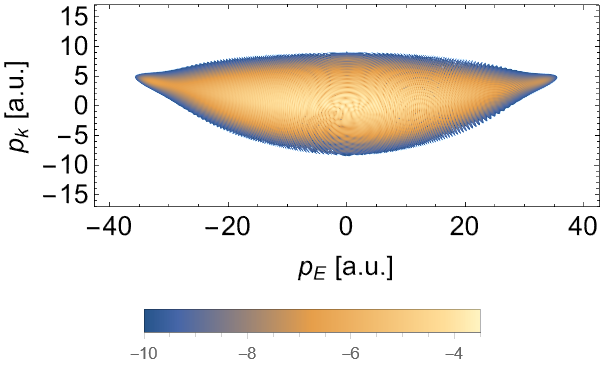}
\includegraphics[width=0.32\textwidth]{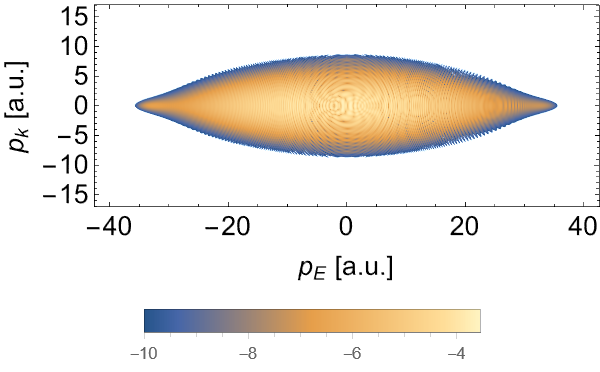}
 \caption{PMD in the case of HECE: (1st column) via relativistic plain SFA (RSFA), (2nd column) via RARM, (3rd) column via nonrelativistic ARM; 
 (first line) for Ar$^{9+}$, $\nu=2.34$, $I_p=17.63$ a.u.,
laser intensity $1.75\times10^{18}$ W/cm$^2$ ($E_0=7.07$ a.u.), and $\omega=0.07$ a.u. ($\upsilon=0.043$, $\xi=0.74$, $Z\omega/E_0=0.14$); (second line) for Ar$^{9+}$,  and XUV beam $\omega=0.5$ a.u. of intensity $1.75\times10^{18}$ W/cm$^2$ ($\upsilon=0.043$, $\xi=0.1$, $Z\omega/E_0=0.99$); (third line) for  Xe$^{36+}$, $\nu=2.626$,  $I_p=93.94$ a.u.,
and X-ray beam of intensity $8.6\times 10^{19}$ W/cm$^2$ ($E_0=65.4$ a.u.) and $\omega=2$ a.u. ($\upsilon=0.1$, $\xi=0.24$, $Z\omega/E_0=1.13$).  }
\label{fig3}
\end{center}
\end{figure*}

\begin{figure}
\begin{center}
\includegraphics[width=0.4\textwidth]{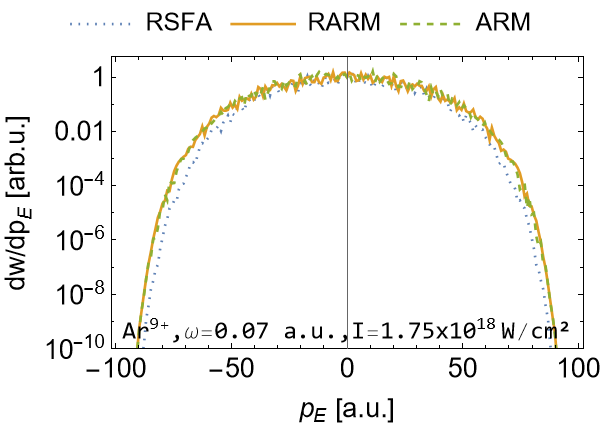}
\includegraphics[width=0.4\textwidth]{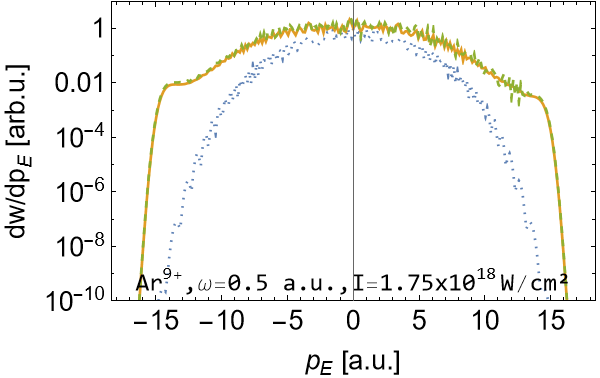}
\includegraphics[width=0.4\textwidth]{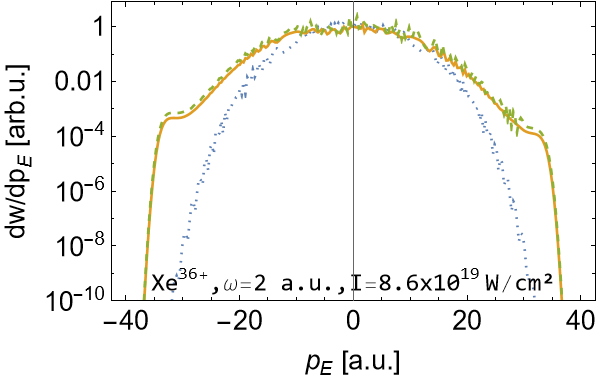}
 \caption{The HECE spectra of Fig.~\ref{fig3} integrated over $p_k$, for the same species and the laser fields:  (blue dotted) via relativistic plain SFA (RSFA), (orange solid) via RARM , (green dashed) via nonrelativistic ARM; (first line) for Ar$^{9+}$, $\nu=2.34$,  $I_p=17.63$ a.u., laser intensity $1.75\times10^{18}$ W/cm$^2$ ($E_0=7.07$ a.u.), and $\omega=0.07$ a.u. ($\upsilon=0.043$, $\xi=0.74$, $Z\omega/E_0=0.14$); (second line) for Ar$^{9+}$,  and XUV beam $\omega=0.5$ a.u. of intensity $1.75\times10^{18}$ W/cm$^2$ ($\upsilon=0.043$, $\xi=0.1$, $Z\omega/E_0=0.99$); (third line) for  Xe$^{36+}$, $\nu=2.626$, $I_p=93.94$ a.u., and X-ray beam of intensity $8.6\times 10^{19}$ W/cm$^2$ ($E_0=65.4$ a.u.) and $\omega=2$ a.u. ($\upsilon=0.1$, $\xi=0.24$, $Z\omega/E_0=1.13$). The distributions are rescaled to the peak value.} 
\label{fig4}
\end{center}
\end{figure}

\section{ Comparison with the nondipole CCSFA }

To test the derived RARM theory, it will be useful to compare its results with those of the nondipole CCSFA describing Coulomb effects in strong field ionization in the nondipole regime. In the nondipole CCSFA only the first relativistic correction to the dipole theory of the order of 1/c  is included. We expect that the fully relativistic theory will coincide with the nondipole one in the limit $I_p/c^2\ll 1$, with significant deviations at $I_p\sim c^2$, as $1/c^2$ terms are neglected in the nondipole theory.  

While in the nonrelativistic theory the electron transverse momentum distribution at the tunnel exit has a peak at zero momentum, in the relativistic treatment the peak is shifted to $p_{k0}=I_p/3c$ along the laser propagation direction due to the sub-barrier effect of the laser magnetic field \cite{Klaiber_2013}. Recently, we showed \cite{He_2022} within nondipole CCSFA that the sub-barrier Coulomb effect increases counter-intuitively the nondipole shift of the longitudinal momentum $p_k$ at the tunnel exit: 
\begin{eqnarray}
p_k&=&p_{k0}+\delta p_k \label{dpk}\\ 
\delta p_k&=& 6(E_0/E_a) p_{k0}. \nonumber
\end{eqnarray}
The CC effect induces an additional dependence of the longitudinal momentum shift on $E_0/E_a$. 
Let us compare the RARM result for the relativistic shift of the peak of the longitudinal momentum due to the sub-barrier CC  with the nondipole theory of \cite{He_2022}, see Fig.~\ref{fig2}, where $\delta p_k/[(E_0/E_a) p_{k0}]$ for hydrogenlike highly charged ion is presented. According to the nondipole approximate theory \cite{He_2022}, $\delta p_k/[(E_0/E_a) p_{k0}]=6$ [Eq.~(\ref{dpk})]. 
The relativistic shift of the peak of the longitudinal momentum presented in Fig.~\ref{fig2} is calculated  analytically  using the exact RARM theory.
To this end,  the atomic as well as the laser action is expanded up to the next to leading order with respect to  $E_0/E_a$:
\begin{eqnarray}
S_0(p_k)&=&S_0(p_{k0})-\frac{2 \kappa\lambda  \left(\lambda ^2+2\right)}{\sqrt{12-3 \lambda ^2} \left(\lambda ^2+1\right)^2}\frac{(p_k-p_{k0})^2}{E_0}.\nonumber\\
S_1(p_k)&=&S_1(p_{k0})+\frac{2 \sqrt{-\lambda ^4+5 \lambda ^2-4} \left(1-\frac{\lambda ^2+\lambda -2}{\lambda }\right)}{(\lambda ^2+1)\kappa}(p-p_{k0}).\nonumber\\
\end{eqnarray}
The longitudinal momentum distribution is given by $\exp\{S\}=\exp[S_0(p_{k})+S_1(p_{k})]$. It has a maximum at $p_k-p_{k0}=-S_1'(p_{k0})/S_0''(p_{k0})$, which reads after using the expansion over $\lambda-1 \sim I_p/c^2$:
\begin{eqnarray}
 p_k-p_{k0}&=&-\frac{\sqrt{3} \left(4-\lambda ^2\right)^{3/2} \left(\lambda ^2-2\right) \left(\lambda ^2+1\right)}{\lambda ^2 \left(\lambda ^2+2\right)}\frac{E_0}{\kappa^3} p_{k0}\label{pkexact}\\
 &\approx & \left[6-28(\lambda-1)\right]\frac{E_0}{\kappa^3} p_{k0}+{\cal O}(\lambda-1) \label{pk000}
\end{eqnarray}
with $p_{k0}=c(\lambda^2-1)/2\lambda\approx I_p/3c$. Thus, the first term of the shift of the most probable momentum in the propagation direction due to the sub-barrier CC  
corresponds to the nondipole result, and the second term $\sim \lambda-1=I_p/(3c^2)$ is the relativistic CC.  The momentum shift coincides with the nondipole result at small $I_p/c^2\ll 1 $. It  is reduced when taking into account the relativistic corrections $\sim I_p/c^2$. This is because the sub-barrier CC originates from the bound state CC, as discussed in \cite{He_2022}. The decrease of the parameter $\nu\approx 1-I_p/c^2$ with higher $I_p/c^2$, see Eq.~(\ref{nunu}), yields larger width of the bound state in momentum space. Then,  the most probable sub-barrier tunneling trajectory begins at the atomic core with larger $p_k $ ending up at the tunnel exit with a smaller one, because the magnetic field-induced momentum drift along the propagation direction is fixed.

Heuristically, the momentum shift can be estimated via $S_a'(p_k)\sim\partial_{p_k}\{\ln[\kappa \sqrt{\mathbf{r}(\eta)^2}]\}\sim\partial_{p_k}\{\ln[\kappa\sqrt{\mathbf{p}^2_i(s-t_i)^2}/\Lambda]\}\sim\partial_{p_k}\{\ln(\kappa^2(s-t_i)/\Lambda)\}\sim\partial_{p_k}\{\ln(\Lambda)\}\sim \partial_{p_k}\{p_k/c\}\sim 1/c$. With $p_{k,0}\sim\kappa^2/(6c)$ and $S_0''(p_{k0})\sim\kappa/E_0$,  the momentum shift $-S_a'(p_{k0})/S_0''(p_{k0})\sim6E_0/\kappa^3p_{k0}$ follows.

\section{High-energy Coulomb enhancement in the case of HCIs}\label{sec:HECEHIC}

We apply the relativistic ARM theory for the investigation of the Coulomb enhancement effect at the cutoff of the direct ionization channel in the relativistic domain with HCIs. This effect of the high-energy Coulomb enhancement (HECE) in the nonrelativistic regime is known, described in \cite{Keil_2016,He_2018}. The effect emerges due to the Coulomb momentum transfer in the continuum. The electron trajectory that ends up at the cutoff of the direct channel starts at the tunnel exit at relatively weak fields and stay near the exit long time, obtaining rather large Coulomb momentum transfer \cite{He_2018}. The parameter which quantifies HECE is $Z\omega/E_0$ \cite{Keil_2016}.
In the calculation of the continuum CC, the continuum action is expanded in $E_0/E_a$, which yields an expansion in the imaginary part of the complex trajectory:
\begin{eqnarray}
S_1(\mathbf{r(\eta)})=S_1({\rm Re}[\mathbf{r}(\eta)])+i {\rm Im}[\mathbf{r}(\eta)]\cdot{\boldsymbol \nabla}S_1({\rm Re}[\mathbf{r}(\eta)]).
\end{eqnarray}

We calculated PMD for three cases via Eq.~(\ref{m1R}), presented in Fig.~\ref{fig3}. In the first case we consider HECE for Ar$^{9+}$ $I_p=479.76$ eV, $Z_{eff}=14.008$, $\nu=2.34$, laser intensity $1.75\times10^{18}$ W/cm$^2$ ($E_0=7.07$ a.u.) using IR laser beam with $\omega=0.07$ a.u. ($\upsilon=0.043$, $\xi=0.74$, $Z\omega/E_0=0.14$). In the second case the same atomic species are used with XUV laser beam ($\omega=0.5$ a.u.) of the same high intensity $1.75\times10^{18}$ W/cm$^2$ ($\upsilon=0.043$, $\xi=0.1$, $Z\omega/E_0=0.99$). And in the third example we consider Xe$^{36+}$ ($I_p=2556$ eV, $\nu=2.626$) exposed to the strong X-ray field ($\omega=2$ a.u.) $E_0=65.4$ a.u. ($\upsilon=0.1$, $\xi=0.24$, $Z\omega/E_0=1.13$). To elucidate the HECE effect we compare PMD via RARM with the plain relativistic SFA. The transverse width of PMD is $p_B=\sqrt{E_0/\kappa}/2$.

In the first example, the continuum relativistic parameter $\xi$ is the largest. Consequently, we see the parabolic dependence of $p_k$ with respect to $p_E$, which is typical for the electron relativistic dynamics in the continuum, and absent in the nonrelativistic consideration (3rd column in Fig.~\ref{fig3}). However, the Coulomb enhancement (HECE) parameter $Z\omega/ E_0$ is the smallest in the first example, and we do not see a significant Coulomb effect, HECE, as the integrated spectrum over $p_k$ coincides with the plain SFA result. The HECE parameter increases for the second and the third cases, which results in appearance of significant shoulders in PMD at $2U_p$ energies. The continuum relativistic features in PMD also enhance. The relativistic and nonrelativistic PMDs via ARM are clearly distinguishable (2nd and 3rd columns) by the parabolic feature in $p_k$ dependence of $p_E$, however, after $p_k$ integration the HECE features are the same (Fig.~\ref{fig4}). The bound state relativistic character is not very pronounced in the given examples as $\upsilon<0.1$.\\

\section{Conclusion}\label{sec:conclusion}

We have generalized ARM theory for the relativistic regime of strong-field ionization. The CCSFA based on the eikonal wave function for the continuum electron (accounting for the Coulomb interaction of the outgoing electron with the atomic core) has a singularity in the eikonal phase at the Coulomb center, where the strong-field ionization starts in the imaginary time. While in the PPT theory the singularity is remedied via matching the continuum wave function to the undisturbed bound state one, in the ARM theory this procedure is equivalent to the shift of the starting point of the time integration in the ionization amplitude by an appropriate imaginary value. In this paper we have found how the value of the corresponding imaginary time shift is modified in the relativistic regime, which eliminate the singularity of the relativistic CCSFA amplitude for ionization.

The advantage of RARM with respect to CCSFA is that it simplifies the calculations of the ionization amplitude using SPA. However, CCSFA offers a possibility for systematic second order Coulomb corrections when using SPA in the coordinate integration, rather than the matching procedure with the bound state. Moreover, CCSFA allows for the development of the generalized eikonal approximation to treat CC at hard recollisions. For sub-barrier CC, the RARM provides results similar to the nondipole CCSFA.

Finally, we employed RARM theory to calculate the Coulomb enhancement of the above-threshold ionization yield at the cutoff of the directly ionized electrons in the relativistic regime.

\bibliography{strong_fields_bibliography}

 \end{document}